# Innovative Weak Formulation for The Landau-Lifshitz-Gilbert Equations

H. Szambolics[1,2], J.-Ch. Toussaint[1,2], L. D. Buda-Prejbeanu[2,3], F. Alouges[4], E. Kritsikis[2,3] and O. Fruchart[1]

[1]Institut Neel, CNRS-UJF, Nanosciences Department, 25 Av. des Martyrs, BP 166, 38042 Grenoble cedex 9, France
[2]Institut National Polytechnique de Grenoble, 46 Av. Félix Viallet, 38031 Grenoble cedex 1, France
[3]Laboratoire SPINTEC, CEA-CNRS-UJF-INPG, Bât. 1005, 17 Rue des Martyrs, 38054 Grenoble cedex 9, France
[4]Département de Mathématiques, Faculté des Sciences d'Orsay, Université Paris-Sud 11, Bât. 425, 91405 Orsay Cedex, France

A non-conventional finite element formalism is proposed to solve the dynamic Landau-Lifshitz-Gilbert micromagnetic equations. Two bidimensional test problems are treated to estimate the validity and the accuracy of this finite element approach.

## I. INTRODUCTION

THE LANDAU-LIFSHITZ-GILBERT (LLG) equation [1] describes the magnetization dynamics of ferromagnetic systems. The equation reads as:

$$\mathbf{v} = -\mu_0 \gamma (\mathbf{m} \times \mathbf{H}_{eff}) + \alpha (\mathbf{m} \times \mathbf{v}), \quad (1)$$

where $\mathbf{m}$ is the local magnetization vector, $\mathbf{v} = \partial \mathbf{m}/\partial t$, $\gamma$ is the gyromagnetic factor, $\alpha$ is the Gilbert damping constant and $\mathbf{H}_{eff}$ is the effective field obtained by variational derivation of the total energy $E_{tot}$ with respect to $\mathbf{m}(\mathbf{r})$ [2]. This field results from the main interactions in a micromagnetic system and can be written as the sum of four fields: exchange field $\mathbf{H}_{ex}$, anisotropy field $\mathbf{H}_{anis}$, magnetostatic field $\mathbf{H}_m$ and applied field $\mathbf{H}_{app}$.

One important hypothesis of micromagnetism is that the local magnetization vector must be constant in magnitude [3]. Thus a constraint

$$g(\mathbf{m}) = 1 - \mathbf{m}^2 = 0 \quad (2)$$

is imposed on the magnetization. Moreover the so-called Brown condition $\partial \mathbf{m}/\partial n = 0$ on the surface $S$ of the magnetic body has to be taken into account.

From a mathematical point of view, the local form (1) with the constraint (2) and the boundary conditions constitute the so-called "strong form" of the problem to solve. An option for solving numerically this kind of problems is the finite element (FE) method [4], [5]. As the FE approach calculates the solution of a "weak formulation", one must transform the strong form into a weak form. The methodology is always the same: the partial differential equation is multiplied by a "test function" and integrated over the domain of calculus. An important point is the choice of the test functions. It will be shown latter that the dynamics of the magnetic system might be altered if not proper test functions are chosen.

## II. MOTIVATION

Usually, the weak form of (1) is obtained by projecting the physical equations onto vector test functions $\boldsymbol{\varphi}$ belonging to the same vector space as the one in which the solution is sought [6], [7]. As a first step we take into account only the exchange field $\mathbf{H}_{ex} = (2A_{ex}/\mu_0 M_S)\Delta \mathbf{m}$. The following weak form for LLG (noted WF1) is obtained:

$$\int_\Omega \boldsymbol{\varphi} \cdot (\mathbf{v} - \alpha(\mathbf{m} \times \mathbf{v})) d\Omega - \frac{2A_{ex}\gamma}{M_S} \sum_l \int_\Omega \left( \mathbf{m} \times \frac{\partial \mathbf{m}}{\partial x_l} \right) \cdot \frac{\partial \boldsymbol{\varphi}}{\partial x_l} d\Omega = 0, \quad (3)$$

$\Omega$ being the magnetic domain, $A_{ex}$ the exchange constant and $M_S$ the spontaneous magnetization. The solution $\mathbf{m}$ of WF1 must satisfy (2), a constraint that is very difficult to deal with [6]. Furthermore, even when a correct method for its treatment is used, unfortunately, because of the interpolation of the magnetization using Lagrange polynomials, this condition is fully respected only at the nodes of the finite element discretization, and only partially in the elements.

To avoid the problems resulting from the constraint Alouges proposed in 2006 an unusual weak formulation for the LLG equations [8]. He considered that due to the constraint (2) the vector fields $(\mathbf{m},\mathbf{v})$ belong to mutually orthogonal subspaces, namely $\mathbf{v}$ is in the tangent space to $\mathbf{m}$ at each mesh node. Based on this, he adapted the projection method by choosing vector test functions $\mathbf{w}$ that also belong to this typical subspace. He replaced thus the "classical" test functions $\boldsymbol{\varphi}$ by $\mathbf{m}\times\mathbf{w}$, such as $\mathbf{m}\cdot\mathbf{w}=0$, without loss of generality. Taking into account these, a new weak form (WF2 hereafter) is obtained from (3), again only for the exchange term:

$$\int_\Omega \mathbf{w} \cdot (\alpha \mathbf{v} + \mathbf{m} \times \mathbf{v}) d\Omega + \frac{2A_{ex}\gamma}{M_S} \int_\Omega \nabla \mathbf{m} \cdot \nabla \mathbf{w} d\Omega = 0. \quad (4)$$

The main advantage of WF2 is that the constraint (2) is implicitly verified because the solution $\mathbf{v}$ is in the tangent subspace to $\mathbf{m}$. Such a problem is easier to treat than (3). Once $\mathbf{v}$ calculated the local magnetization vector can be easily reconstructed.

The estimation of the integral in the exchange term is more sensitive to interpolation errors as the number of terms in **m** and ∇**m** is large. As in WF1 the exchange term contains both **m** and its space derivative, whereas in WF2 only ∇**m** occurs, WF1 is more sensitive to this kind of errors than WF2.

We implemented and tested both weak formulations. In the following we present the results for a first 2D test case consisting of a 2x2nm square, in which the magnetic moments are coupled by exchange. A sinusoidal magnetic distribution is imposed at the beginning of the simulation. At equilibrium all the moments should be aligned.

The dynamics of the magnetization was calculated with WF1 and WF2. The relaxation process towards equilibrium calculated by WF1 and WF2 is compared with the one obtained by a finite difference (FD) approach, earlier implemented in the GL_FFT software (by J.C. Toussaint, © Institut Néel) [9]. The value of the damping parameter was taken to be 0.02.

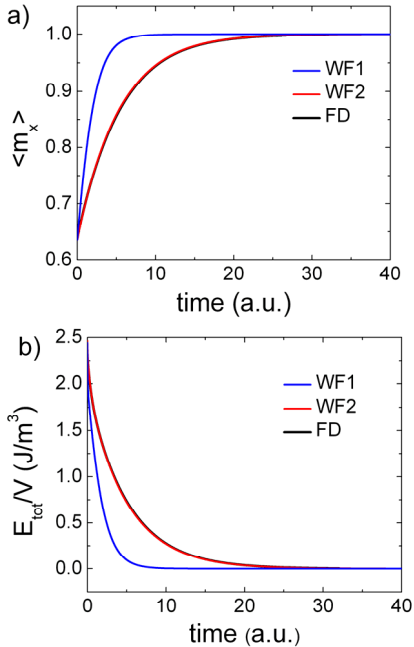

Fig. 1. Evolution of a) the average value of the $m_x$ magnetization component and b) the energy density as a function of time. The time is expressed in arbitrary units. The evolution of the FD and the WF2 curves are quite similar, for both $m_x$ and $E_{tot}$, whereas the blue curve corresponding to WF1 is clearly over-damped.

The comparison (Fig. 1) shows that although the equilibrium states are the same, the paths followed by the magnetization in the relaxation process are quite different. Using WF1, one notices a faster decrease of the energy than with the WF2 approach and the FD approach. The WF1 motion is artificially over-damped, most likely because of the treatment of the constraint (2).

### III. WEAK FORMULATION FOR MICROMAGNETISM

The results for the first test case encouraged us to implement a FE formulation based on Alouges's idea including all four field terms. Thus a complete weak formulation is obtained:

$$\alpha \int_\Omega \mathbf{w} \cdot \mathbf{v}\, d\Omega + \int_\Omega \mathbf{w} \cdot \mathbf{m} \times \mathbf{v}\, d\Omega \\ + \frac{2A_{ex}\gamma}{M_S} \int_\Omega \nabla \mathbf{w} \cdot \nabla \mathbf{m}\, d\Omega - \mu_0 \gamma \int_\Omega \mathbf{w} \cdot \mathbf{H}\{\mathbf{m}\}\, d\Omega = 0. \quad (5)$$

Here **H**{**m**} stands for the sum of the magnetostatic, applied and anisotropy field. There are two possible ways of finding the magnetostatic field in FE: by using either a magnetic vector potential approach including the Coulomb gauge, or more simply, using a magnetic scalar potential approach. In both cases the regularity of the magnetic potential at infinity is imposed [10]. To treat this, a spatial transformation [11] is used, converting the infinite exterior domain into a finite one [6].

**v** is expressed as a finite difference of the solutions at times n and n+1, thus the magnetization vector at time n+1 is obtained by

$$\mathbf{m}^{n+1} = \mathbf{m}^n + \delta t\, \mathbf{v} \quad (6)$$

and must be normalized at each mesh node.

#### A. Time Integration Scheme

An explicit Euler scheme requires very small time steps. In a FE approach, due to the non-homogeneity of the spatial discretization, the time step is bounded by the size of the smallest mesh element, leading thus to its dramatic reduction, and thus decreasing the efficiency of the method. To avoid it we adopted a classical θ-scheme [12]. The magnetization in the exchange term has been expressed as $\mathbf{m} + \theta \delta t\, \mathbf{v}$. Consequently the weak formulation (5) is modified as follows:

$$\int_\Omega \mathbf{w} \cdot (\alpha \mathbf{v} + \mathbf{m} \times \mathbf{v})\, d\Omega + \theta \delta t \frac{2A_{ex}\gamma}{M_S} \int_\Omega \nabla \mathbf{w} \cdot \nabla \mathbf{v}\, d\Omega \\ + \frac{2A_{ex}\gamma}{M_S} \int_\Omega \nabla \mathbf{w} \cdot \nabla \mathbf{m}\, d\Omega - \mu_0 \gamma \int_\Omega \mathbf{w} \cdot \mathbf{H}\{\mathbf{m}\}\, d\Omega = 0. \quad (7)$$

where 0≤θ≤1. In particular, for θ=0 one retrieves an explicit scheme, for θ=1/2 a Crank-Nicholson-like scheme, and finally θ=1 represents an implicit integration scheme. In the simulations we used θ=1/2 because its accuracy is better than for the implicit one (θ=1), although is not of order 2 in time due to normalization.

As a physical system relaxing towards equilibrium loses energy, one should verify that the above mentioned integration scheme indeed describes a dissipative process. In other terms, one has to check whether the energy difference $\Delta E_n$ between two consecutive time steps is negative.

$$\Delta E_n = E\{\mathbf{m}^{n+1}\} - E\{\mathbf{m}^n\} = E\{\mathbf{m}^n + \delta t\, \mathbf{v}\} - E\{\mathbf{m}^n\} \quad (8)$$

For sake of simplicity, only the exchange term is kept in the following proof. In this context, the system's energy reads as $E\{\mathbf{m}^n\} = \int A_{ex} (\nabla \mathbf{m}^n)^2$ at each time step n.

By the substitution of **w** by **v** in (7), a mathematically valid operation as they belong to the same subspace, we get the following property:

$$\alpha \int_\Omega \mathbf{v}^2 d\Omega + \theta\, \delta t \frac{2A_{ex}\gamma}{M_S} \int_\Omega (\nabla \mathbf{v})^2 d\Omega = -\frac{2A_{ex}\gamma}{M_S} \int_\Omega \nabla \mathbf{v} \cdot \nabla \mathbf{m}^n d\Omega \quad (9)$$

By introducing this in (8), the expression of $\Delta E_n$ is obtained:

$$\Delta E_n = -\alpha\, \delta t \int_\Omega \mathbf{v}^2 d\Omega - \left(\theta - \frac{1}{2}\right) \delta t^2 \frac{2A_{ex}\gamma}{M_S} \int_\Omega (\nabla \mathbf{v})^2 d\Omega. \quad (10)$$

It is readily seen that by choosing $\theta \in [1/2, 1]$ the system's energy is guaranteed to decrease in time. A general proof taking into account all the field terms was established and will be published elsewhere.

## IV. RESULTS

In thin films with a perpendicular magnetocrystalline anisotropy of moderate strength (FePd alloys, Co(0001) or Co/Pt multilayers) particular equilibrium configurations at remanence are observed, made of a periodic modulation of the perpendicular component of the magnetization that leads to parallel stripe domains [13]. Such magnetic bodies are thus well adapted to 2D micromagnetic simulations since the magnetization is nearly invariant along the stripes' direction (Oz axis) and is periodic in the other in-plane direction (Ox axis).

Because of the periodic character of the structure the simulated system can be reduced to only a region that has the length equal to the equilibrium period. The equilibrium period is determined by the material parameters describing the system. The material parameters used in our simulations are: $A_{ex}=2\cdot 10^{-11}$ J/m, $\mu_0 M_S=1$ T, $K_{anis}=5\cdot 10^5$ J/m$^3$. Using these values, the equilibrium period of the magnetic system was determined to be around 200nm. Thus the simulated geometry consists of a rectangular body with the length of 200nm and height of 40nm. A schematic representation of the model system is given in Fig. 2:

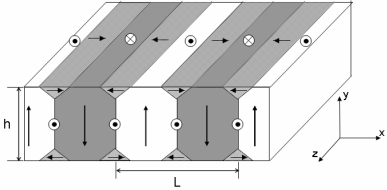

Fig. 2. Schematic representation of the stripe domain structure. The symmetry properties permit to reduce the size of the simulated system to only a unit, of height h=40nm and length L=200nm.

The magnetization dynamics towards equilibrium was calculated with several values of the damping parameter by using the WF2 based approach and compared to those obtained with the FD software GL_FFT. We present hereafter only the study realized with the last value of the damping parameter (0.03) because of its relevance in testing the accuracy of the WF2 implementation.

As initial magnetization configuration a sinusoidal profile was chosen:

$$m_x = 0, \quad m_y = \cos\left(2\pi \frac{x}{L}\right), \quad m_z = \left|\sin\left(2\pi \frac{x}{L}\right)\right|. \quad (11)$$

Figure 3a) shows the time evolution of the magnetization component oriented along the Oz direction. The oscillations are related to the value of the damping parameter ($\alpha$=0.03), and indicate a highly non-linear dynamic behavior.

In Fig. 3b) the time evolution of the total energy is shown, consistent with a dissipation process towards equilibrium. There is a small energy difference around 1% between the equilibrium values. The residual discrepancy may be attributed to the different ways to evaluate the total energy: FD uses local estimations of the magnetization vector and the effective field, whereas in FE the energy expression

$$E_{tot} = \int_\Omega A_{ex} [\nabla \mathbf{m}]^2 d\Omega + \int_\Omega K_{anis}\left[1-(\mathbf{u_K} \cdot \mathbf{m})^2\right] d\Omega \\ -\int_\Omega \mu_0 M_s\, \mathbf{m} \cdot \mathbf{H}_{app} d\Omega - \int_\Omega \frac{1}{2}\mu_0 M_s\, \mathbf{m} \cdot \mathbf{H_m} d\Omega \quad (12)$$

is applied to the magnetization field interpolated on each element.

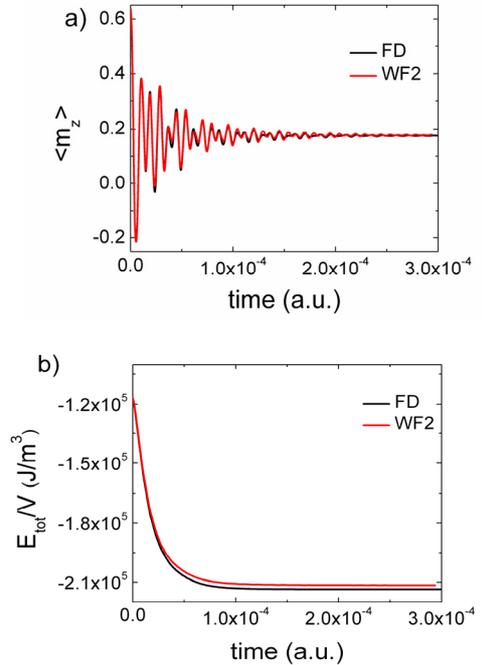

Fig. 3. Temporal evolution of a) the average value of the $m_z$ magnetization component and b) the total energy density. The results issued by the GL_FFT software and the WF2 approach are in very good agreement. For the equilibrium energy value there is a difference of 1% between the results. For the magnetization component, even though only the first oscillations are reproduced exactly by the WF2 approach, the characteristic times are the same.

Several intermediate configurations are represented in Fig. 4 to illustrate the good agreement between FD and WF2. In the first part of the motion, where the oscillation amplitude of $m_z$ is very large the domain walls are quite distorted. As the equilibrium approaches the final structure of the vortex walls is formed and they oscillate around the equilibrium position until the system stabilizes.

The last configuration represents the equilibrium state. The magnetization mainly lies in the Oxy plane, forming the magnetic domains parallel to the easy axis. As expected from the value of the quality factor (1.25) a complex wall structure is formed: in the center the domain wall corresponds to a vortex structure, whereas near the surfaces two flux closure domains with opposite magnetization orientation appear. The walls are separated by a distance of L/2. The maximum misalignment between the FD and FE equilibrium configurations was determined. The most important discrepancies lie in the region where the domain walls are placed, but overall, the angle between the magnetization calculated with the two approaches does not exceed 0.05 rad.

It is important to notice that for small damping values ($\alpha \approx 5 \cdot 10^{-2}$) the relaxation process is quite complex. Sometimes the domain walls can become so distorted that, if not proper values for the space discretization are used, unphysical processes might occur. As a rule of thumb, in micromagnetic simulations the maximum mesh size must be smaller than the minimum of any of the characteristic lengths: the exchange and the Bloch length [2]. In our case the exchange length is $l_{ex}$=5.01 nm and the Bloch length $l_B$=6.32 nm. Taking these into consideration, the simulations have been carried out using a mesh size of 2 nm, thus smaller than the value given by the rule of thumb. This however is required because for example in the third configuration in Fig. 4, the vortex core extends only over 4 nm. In this case a mesh size of 5 nm would not be recommended in order to prevent the non-physical reversal of the vortices.

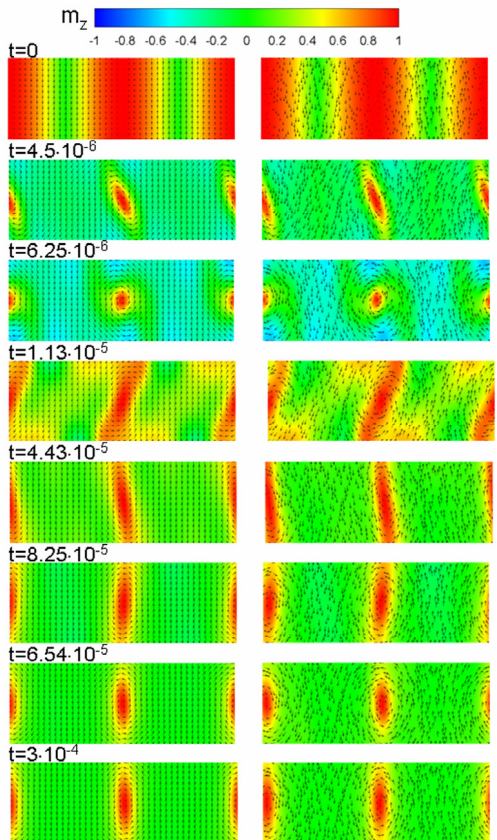

Fig. 4. Magnetization distribution calculated by FD (left) and WF2 (right) at several time steps (the time being expressed in arbitrary units). The $m_x$ and $m_y$ magnetization components are represented by arrows and the $m_z$ component by a color scale.

## V. Conclusions

This study emphasizes that the essential point in the micromagnetic modeling by finite elements is not the one related to meshing, but the problem of implementing correct weak formulations.

We proposed and compared two weak formulations, WF1 obtained by direct projection of the LLG equations and WF2 obtained by a more sophisticated method. The property of the integration scheme to be dissipative was demonstrated only for the latter one. We showed, for a simplified test case that WF1 is intrinsically more dissipative that WF2, and thus is not adapted for the description of magnetization dynamics for realistic damping values ($\alpha$=$10^{-2}$).

In the case of a 2D stripe system the WF2 implementation was shown to reproduce with high accuracy the magnetization dynamics obtained with a finite difference approach. A good agreement was obtained for averaged quantities, like the magnetization and the total energy as a function of time, and also for local ones, like magnetization vector distributions.

### Acknowledgment

This work was supported by the French research cluster «Microélectronique, Nanosciences et Nanotechnologies» of the Rhône-Alpes region.

### References

[1] T. L. Gilbert, "A phenomenological theory of damping in ferromagnetic materials," IEEE Trans. Magn., vol. 40, pp. 3443-3449, Nov. 2004.
[2] A. Hubert and R. Schafer, *Magnetic Domains*, Springer, New York, 1998.
[3] W. F. Brown Jr., *Micromagnetics*, Interscience Publishers, J. Wiley and Sons, New York, 1963.
[4] D. Braess, *Finite Elements*, 2nd ed., Cambridge University Press, Cambridge, 2001.
[5] F. Vermolen, *Introduction into Finite Elements*, http://ta.twi.tudelft.nl/users/vermolen/wi3098/wi3098.pdf
[6] H. Szambolics, L. D. Buda-Prejbeanu, J.-Ch. Toussaint and O. Fruchart, "A constrained finite element formulation for the Landau-Lifshitz-Gilbert equations", submitted for publication.
[7] B. Yang and D. R. Fredkin, "Dynamical micromagnetics by the finite element method", IEEE Trans. Magn., vol. 34, pp. 3842-3852, Nov. 1998.
[8] F. Alouges and P. Jaisson, "Convergence of a finite elements discretization for Landau-Lifshitz equations", Math. Mod. Meth. Appl. Sci, vol. 16, pp. 299-313, 2006.
[9] B. M. Kevorkian, *Contribution à la modélisation du retournement d'aimantation. Application a des systèmes magnétiques nanostructures ou de dimensions réduites*, Ph.D. dissertation, Université Joseph Fourier, Grenoble, 1998.
[10] J. D. Jackson, *Classical Electrodynamics*, 3rd ed., Wiley, New York, 1999.
[11] X. Brunotte, G. Meunier and J.-F. Imhoff, "Finite modeling of unbounded problems using transformations", IEEE Trans. Magn., vol 28, pp. 1663-1666, 1992.
[12] B. Lucquin and O. Pironneau, *Introduction au Calcul Scientifique*, Masson, Paris, 1995.
[13] J.-Ch. Toussaint, A. Marty, N. Vukadinovic, J. Ben Youssef and M. Labrune, "A new technique for ferromagnetic resonance calculations," Comp. Mat. Sci, vol 24, pp. 175-180, May 2002.